\documentclass[conference]{IEEEtran}
\IEEEoverridecommandlockouts
\usepackage{array}
\usepackage{multirow}
\usepackage{booktabs}
\usepackage{cite}
\usepackage{amsmath,amssymb,amsfonts}
\usepackage{algorithmic}
\usepackage{graphicx}
\usepackage{textcomp}
\usepackage{xcolor}
\usepackage{diagbox}
\def\BibTeX{{\rm B\kern-.05em{\sc i\kern-.025em b}\kern-.08em
    T\kern-.1667em\lower.7ex\hbox{E}\kern-.125emX}}
\begin{document}

\title{A Systems Thinking for Cybersecurity Modeling\\
}

\author{\IEEEauthorblockN{Dingyu Yan}
 \IEEEauthorblockA{State Key Laboratory of Information Security, Institute of Information Engineering,\\Chinese Academy of Sciences, Beijing, China\\
 School of Cyber Security, University of Chinese Academy of Sciences, Beijing, China\\
 yandingyu@iie.ac.cn}
}
\maketitle

\begin{abstract}
Solving cybersecurity issues requires a holistic understanding of components, factors, structures and their interactions in cyberspace, but conventional modeling approaches view the field of cybersecurity by their boundaries so that we are still not clear to cybersecurity and its changes. In this paper, we attempt to discuss the application of systems thinking approaches to cybersecurity modeling. This paper reviews the systems thinking approaches and provides the systems theories and methods for tackling cybersecurity challenges, regarding relevant fields, associated impact factors and their interactions. Moreover, an illustrative example of systems thinking frameworks for cybersecurity modeling is developed to help broaden the mind in methodology, theory, technology and practice. This article concludes that systems thinking can be considered as one of the powerful tools of cybersecurity modeling to find, characterize, understand, evaluate and predict cybersecurity.
\end{abstract}


\section{Introduction}
Ever since the concept of "cyberspace" is defined clearly, the boundary of security is extended to the real-world domain related to digital technology rather than the only virtual environment created by computer networks \cite{craigen2014defining}. Though there are already several works dedicated to cybersecurity modeling \cite{al2016cyber} \cite{koutsoukos2017sure}, we still have a vague understanding on cybersecurity. Firstly, it is difficult to exploit and evaluate the synergies among the defensive measures to enhance cybersecurity \cite{xi2015improved}. Despite the large investments in the security field from nation, enterprises and individuals, we cannot know whether these defensive measures can really work against cyberattacks. Secondly, the security system lacks the capacity to measure its current security situation comprehensively and precisely. Nowadays, there is no set of the unified and accepted evaluation system and metrics for cybersecurity modeling. Thirdly, systemic components, factors and their interaction are often ignored and omitted in several models \cite{herley2017sok}. The multiple relationships and interaction among the components greatly increase the difficulty of cybersecurity modeling and analysis.

To better understand the essential characters of cybersecurity and resolve the cybersecurity challenges, several studies attempt to apply systems thinking approaches to the cybersecurity field \cite{young2013systems}\cite{salim2014cyber}. Systems thinking is considered to offer a novel and comprehensive perspective to reveal the entire process of cybersecurity as a system. Also, by these systems thinking approaches, researchers plan to establish a conceptual framework for measuring and evaluating the cybersecurity system and its constitutes \cite{national2017foundational}\cite{spring2017practicing}, such as defensive measures, human factors, security policy. Thus, the typical goals of the systems thinking for cybersecurity modeling are exemplified as follows: (1) discovering the multiple impact factors and their interacting effects; (2) investigating fundamental laws in cybersecurity; (3) exploring the theoretical and real solution to the specific security issue; (4) evaluating effective attack weapons and defense measures in the specific scenario.

This article mainly urges the importance of systems thinking in cybersecurity modeling. Firstly, we analyze the primary characteristics of cybersecurity and the challenges in cybersecurity modeling in Section II. Then, Section III introduces systems thinking and explores how systems thinking is applied for cybersecurity modeling. Finally, we give an example of systems frameworks for cybersecurity modeling in Section IV.
\begin{figure*}
 \label{Fig.1}
 \centering
 \begin{minipage}{16cm}
  \centering
  \includegraphics[width=1\textwidth]{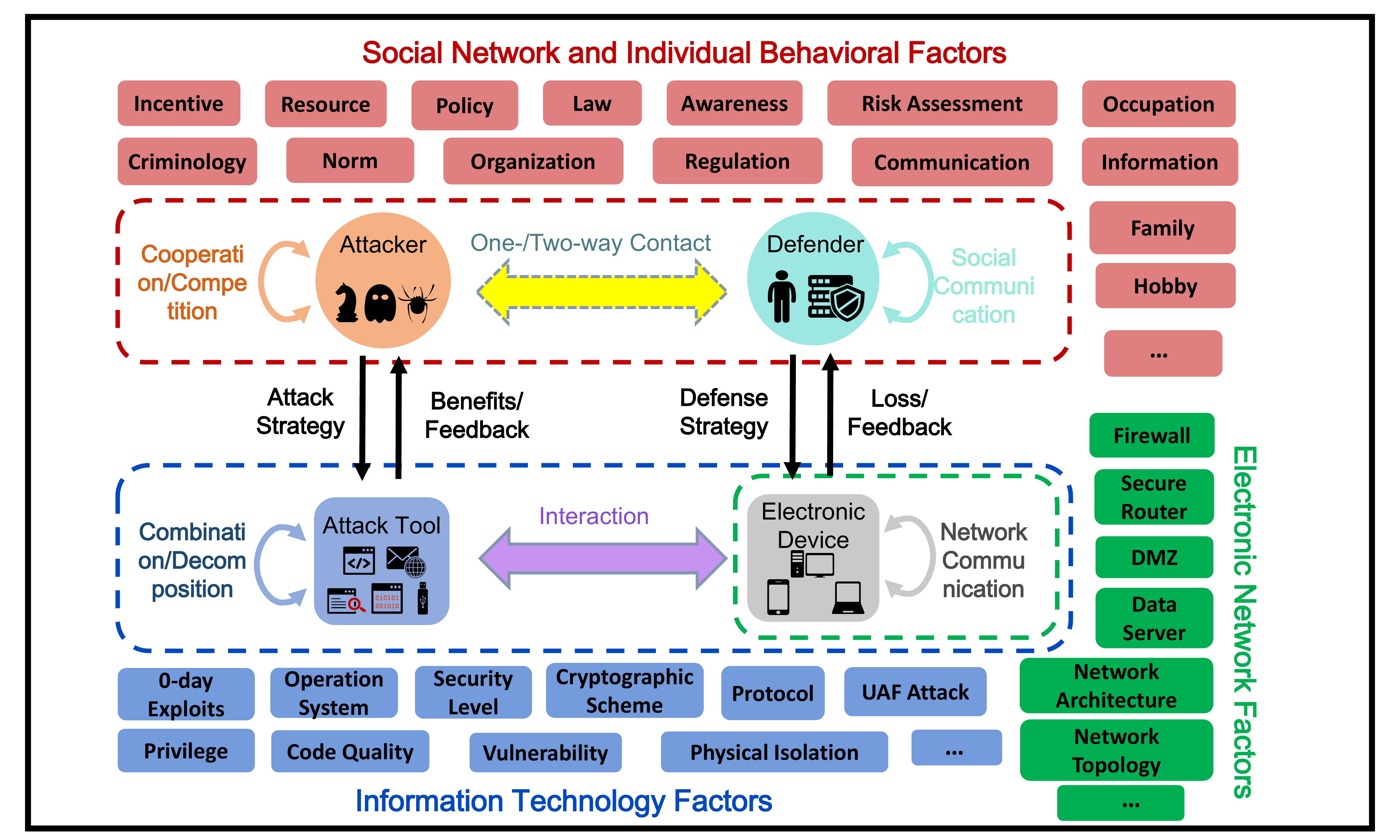}  
 \end{minipage}
 \centering
 \caption{The main components, impact factors and their interaction in Cybersecurity.}
\end{figure*}

\section{Cybersecurity as a Complex System}
\subsection{Characteristics of Cybersecurity}
Cyberspace can be considered as the ultimate complex adaptive system of interconnected heterogeneous components \cite{phister2011cyberspace} \cite{rzeszutko2015insights}, such as multiple types of networks, devices and stakeholders, intertwined with human behavioral and technical factors, as shown in schematic figure Fig.1. In modeling this complex cybersecurity landscape, the four following characteristics are inevitable.

\textbf{Complexity.} Complexity is the most prominent characteristic of the cybersecurity and has infiltrated each part of the cybersecurity landscape \cite{dunlavy2009mathematical}\cite{armstrong2009complexity}. First of all, complexity in cybersecurity embodies the diversity in security issues and the multiplicity in influence factors. Cybersecurity is a complex intercross area, covering multiple fields, such as society, economics, politics, information technology, etc. The cybersecurity issues are stemmed from these fields and are affected by the combination of factors. For example, Flame, an example of the advanced persistent threat attack targeted Middle Eastern countries, is considered as the highly sophisticated and well-planned nation-state cyberattack for military and political motives. Moreover, the interrelationships between components in cyberspace are extremely complicated. Each component could interact with others in each field. Especially, as the center of cyberspace, the human is the interface between the natural environment, human society and information technology. The components of these three fields such as social distance, network architecture clearly alter the individual behavior and strategy, but, in turn, the participant can influence the dynamic change of components.

\textbf{Unpredictability.} As interactions exist among components and joint effects of the multiple types of factors, the cybersecurity as a whole exhibits an unpredictability and complexity \cite{rzeszutko2015insights}. First, the behavior, action and strategy of participants in cyberspace, either adversaries or users, are irrational, unpredictable and nonuniform \cite{oltramari2015towards}. Notably, smart hackers hide themselves by abandoning the conventional attack technology and method, so they are challenging to attribute definitively. Second, vulnerabilities and malfunctions in system and protocol, sometimes, are imperceptible. This feature makes it difficult for the defender to evaluate the system security and analyze the defense effectiveness quantitatively. Third, the cybersecurity system can exhibit the unpredictable emergence \cite{xu2015emergent}. As defined in terms of the system-level patterns, emergence in cybersecurity refers to the new property and macrocosmic phenomenon as a result of the interactions of components in the microcosmic level. Thus, it is difficult for us to evaluate the effectiveness of some specific attack and defense technologies on the whole cyberspace.

\textbf{Dynamics.} To further understand the issues associated with cybersecurity, one must be knowledgeable about the evolution of each component \cite{xu2015cybersecurity}. On the one hand, the state of every component and each interaction between every two components changes dynamically over time. Especially for participants in cyberspace, their behavior and strategy may be dynamical and inconsistent \cite{oltramari2015towards}. On the other hand, the dynamics of cybersecurity is the prerequisite of system emergence\cite{yan2019dynamical}. The same input and environmental conditions do not always guarantee the same output.

\textbf{Asymmetry.} In cybersecurity, there always exists an asymmetry between the attacker and the defender \cite{geers2010challenge}. This asymmetry is presented in the following three aspects. First, the attacker is positive and proactive, while the defender is passive and reactive \cite{Cai2016Moving}. In general, the attacker makes enough preparations in advance, such as vulnerability scanning, intelligence gathering, weaponization, which are ensured not clear to the defender. Moreover, the defender must protect all possible points of the protected object at any time, while the attacker can break through the meticulous defense disposition just by one valid vulnerability. Second, the defender's evaluation of his defense effectiveness is often faulty. As mentioned above, because it is difficult to estimate the effects of the specific defense technology or method on the whole cybersecurity, the defender fails to measure the defense effectiveness comprehensively and accurately. Third, the cost of one attack is less than that of defense. The defend requires an enormous investment of money, labor and resource, regardless of researching new defense technology or establishing the early warning mechanism for large-scale cyber attacks. However, these defense methods and technologies cannot guarantee to resist the cyber attack completely. 

\begin{table*}[!t]
	\renewcommand\arraystretch{1.3}
	\centering
	\caption{Systems Thinking Theories and Methods}
	\label{Table.1}
	\begin{tabular}{p{2cm}<{\centering}p{2cm}<{\centering}p{11.5cm}}
		\multicolumn{3}{c}{Theories}\\
		\midrule
		{Item} &{Key Research} & \multicolumn{1}{c}{Description}\\
		\midrule
		{Systems Theory} &{\cite{salim2014cyber}\cite{tisdale2015cybersecurity}} & {Systems Theory is an interdisciplinary methodology which employs several systems approaches to investigating the systems structure, understanding the complex phenomenon and solving the relevant problems.}\\
		{Game Theory} &{\cite{manshaei2013game}\cite{do2017game}} & {Game Theory attempts to explain the interacting strategy of the players with respect to the utilities of other rational players. In the security game model, the attacker and defender act as the players in the game theory.}\\
		{Cybernetics} &{\cite{adams2013application}\cite{vinnakota2013cybernetics}} & {Cybernetics is a broad study of both living and non-living systems guided by principles of feedback, control, and communications.}\\
		{Catastrophe Theory} &{\diagbox[width=1.2cm,font=\footnotesize\itshape,innerleftsep=-0.2cm,innerrightsep=-0.1cm]{}{}} & {Catastrophe Theory is a mathematical theory for explaining the abrupt changes and discontinuities of state (E.g., server crash, defense failure).}\\
		{Behaviorism} &{\cite{wiederhold2014role}\cite{marble2015human}\cite{anwar2017gender}} & {Behaviorism is a learning approach which focuses on the human behavior in cyberspace. }\\
		\midrule
		\multicolumn{3}{c}{Methods}\\
		\midrule
		{Item} &{Key Research} & \multicolumn{1}{c}{Description}\\
		\midrule
		{Dynamics} &{\cite{xu2015cybersecurity}\cite{zheng2015active}} & {Dynamics is a system methodology technique to model the system problems by dealing with stocks, follows and feedback loops that affect the behavior of the entire system over time.}\\
		{Network Analysis} &{\cite{la2014role}\cite{pino2014network}} & {Network analysis is both a theoretical approach and methodological tool for understanding the interactions between the actors, exploring the network structure effects and studying the relevant factors.}\\
		{Agent-based Model} &{\cite{such2016intelligent}} & {Agent-based model is a way to model or simulate the complex system constituted by autonomous, interacting agents (e.g., individual, group). The heterogeneity in agents' strategy decision and complicated interactions between agents can result in the unpredicted results of the system as whole.}\\
		\bottomrule
	\end{tabular}
\end{table*}

\subsection{Challenges of Cybersecurity Modeling}
At present, cybersecurity modeling is still in its infancy. The existing models and methods are limited to the technical security study, aiming to address the specific technical problem by the technology and approaches \cite{white2017computer}. For example, the modern cryptographic scheme is to solve the problem of preventing the malicious party from obtaining private information. Confidentiality, integrity and availability are the core aspects of the cryptography \cite{stallings2017cryptography}. Based on where the security technique works, it often classifies these technical studies into three classes: applications-based, hosts-based and network-based security technical studies. For instance, code injection is the applications-based study and firewall belongs to both network-based and host-based security study. Though abundant technical works have made significant contributions to the research field of information security, there are still several typical challenges technical study: (1) the formalized description of the technical problems in cybersecurity; (2) the unified pattern of the quantitative and qualitative analysis towards the security technology; (3) coupling the theoretical guidance with the security technology and practice.

Generally speaking, cybersecurity study should involve the all relevant factors in the fields of politics, society, economy and culture, covering the theory, technology and practices. Fig.1 demonstrates the cybersecurity is rather a complex system, where the multiple components, factors and environment are interacted and twisted. Thus, cybersecurity modeling is broader than mere technical study. Recently, researchers have become interested in the human behavior factors in cybersecurity. The Data Breach Investigations Report (DBIR) from Verizon \cite{enterprise2018} represented that human factor continues to be a major issue accounting for the most incidents in enterprise. In cybersecurity, humans play as both developers and users for the security products; act as both adversaries and victims for the cyber attack-defense \cite{yan2018characterizing}. For example, game theory, such as the static game, dynamic game, Bayesian game, is often applied to investigate the interaction between the attacker and defender \cite{manshaei2013game}\cite{do2017game}. In the Stackelberg Security Game (SSG), where a leader makes a decision first and then a follower reacts subject to the leader's action, the attacker acts as the follower, and the defender acts the role of the leader. However, there are still many theoretical and technical difficulties that need to be tackled in characterizing individual behaviors in cybersecurity. For example, human cognitive bias, gambler psychology, and the heterogeneity, dynamics and uncertainty in individual strategy decision can make the conventional method difficult, even invalid, for investigating the role of human factors in cybersecurity \cite{oltramari2015towards}.

\begin{table*}[t]
 \renewcommand\arraystretch{1.5}
 \centering
 \caption{Relevant Stakeholders in Cybersecurity}
 \label{Table.2}
 \begin{tabular}{p{2.5cm}<{\centering}p{5cm}<{\centering}p{8.5cm}}
  \toprule
  
  {Stakeholders } &{clusters} & \multicolumn{1}{c}{Constituent Sub-system Description}\\
  \midrule
  {Government} &{Ministries, law enforcements, regulatory agencies} & {As policy system from the perspective of government}\\
  {Academia} &{Universities, research institutes} & {As research system from the perspective of security researchers}\\
  {Private Sectors} &{Information security enterprises, computer and network companies} & {As market system from the perspective of providers of security services and products}\\
  {Infrastructure} &{Internet service providers, urban managers} & {As urban manage system from the infrastructure managers and planners}\\
  
  \bottomrule
 \end{tabular}
\end{table*}

The cybersecurity study also needs to address the cyber-physical security issues, such as industry control systems, laws and regulations, cybercrime. Playing an essential role in financial services, power grid, transportation and medical system, industry control systems are often selected as targeted for cyberattacks \cite{mclaughlin2016cybersecurity}, especially advanced persistent threat attacks. These cyberattacks tend to disrupt the order of the nation, cause public panic and disorder \cite{yan2019modeling}. For example, Stuxnet, a sophisticated malware with four 0-day exploits targeting the Windows system and one targeting SCADA, delayed the process of Iranian nuclear program \cite{ask2013advanced}. In December 2015, a cyberattack on Ukraine power grid by Black Energy group took place, and about 225 thousand customers lose power before Christmas \cite{Liang2016The}. In order to the fields as mentioned above, other specialists strive for a systematic set of cybersecurity metrics to define, measure and quantify cybersecurity \cite{xi2015improved}\cite{Pendleton2016A}.

\section{Apply Systems Thinking to Cybersecurity Modeling}
\subsection{What is Systems Thinking}
System thinking is a holistic approach intended to analyze how the parts of the system interact and how the emergence changes as a whole entity \cite{arnold2015definition}. Unlike the reductionist thinking, which actually treats the world from a static, simple and one-sided perspective, this holistic thinking emphasizes the complexity, dynamism and entirety of the system, as well as the interconnected and multifaceted relationships between the system components \cite{de2009systems} \cite{xia2017systems}. Systems thinking arose in the early 20th century and now has been used to the diverse research fields, such as public health, environmental protection, urban management and international relationship. Nowadays, a tiny amount of researchers attempt to implement this systems thinking to the cybersecurity study \cite{young2013systems}\cite{salim2014cyber}.

In our opinion, the best study for finest and resilient cybersecurity modeling needs to consider the systems thinking approaches at this stage. On the one hand, systems thinking for cybersecurity does not only treat a particular area of cyberspace, but allows for the cybersecurity of the whole entity. This holistic approach to cybersecurity is more readily able to identify and understand the cybersecurity system, describe the interaction among cyberspace components, predict the evolution of cybersecurity actually and help us address the cybersecurity issues effectively \cite{young2013systems}. Systems thinking helps broaden the cybersecurity study scope to integrate people, environment, government and other vital aspects. On the other hand, unlike the traditional enumerative and analytical methods, which focus the linear and static causality from an individual perspective, systems thinking emphasizes on the complexity in the interaction of constituents of the cyber system. Despite conventional approaches have made significant achievements in network security technology and cryptology, e.g., detection \& prevention technology and public key cryptography \cite{stallings2017cryptography}, these traditional approaches are not enough for us to depict, characterize and predict the cybersecurity issue and its evolution. In this systems perspective, the purpose of cybersecurity modeling is to promote the whole security situation of cyberspace rather just deal with a specific technological challenge. This requires us to apply the systems thinking to gain insight into cybersecurity from a holistic perspective, rather complement the conventional approach in some deficiencies. Thus, this radical shift in cybersecurity modeling is requisite.

\begin{figure*}[t]
 \label{Fig.2}
 \centering
 \begin{minipage}{16cm}
  \centering
  \includegraphics[width=1\textwidth]{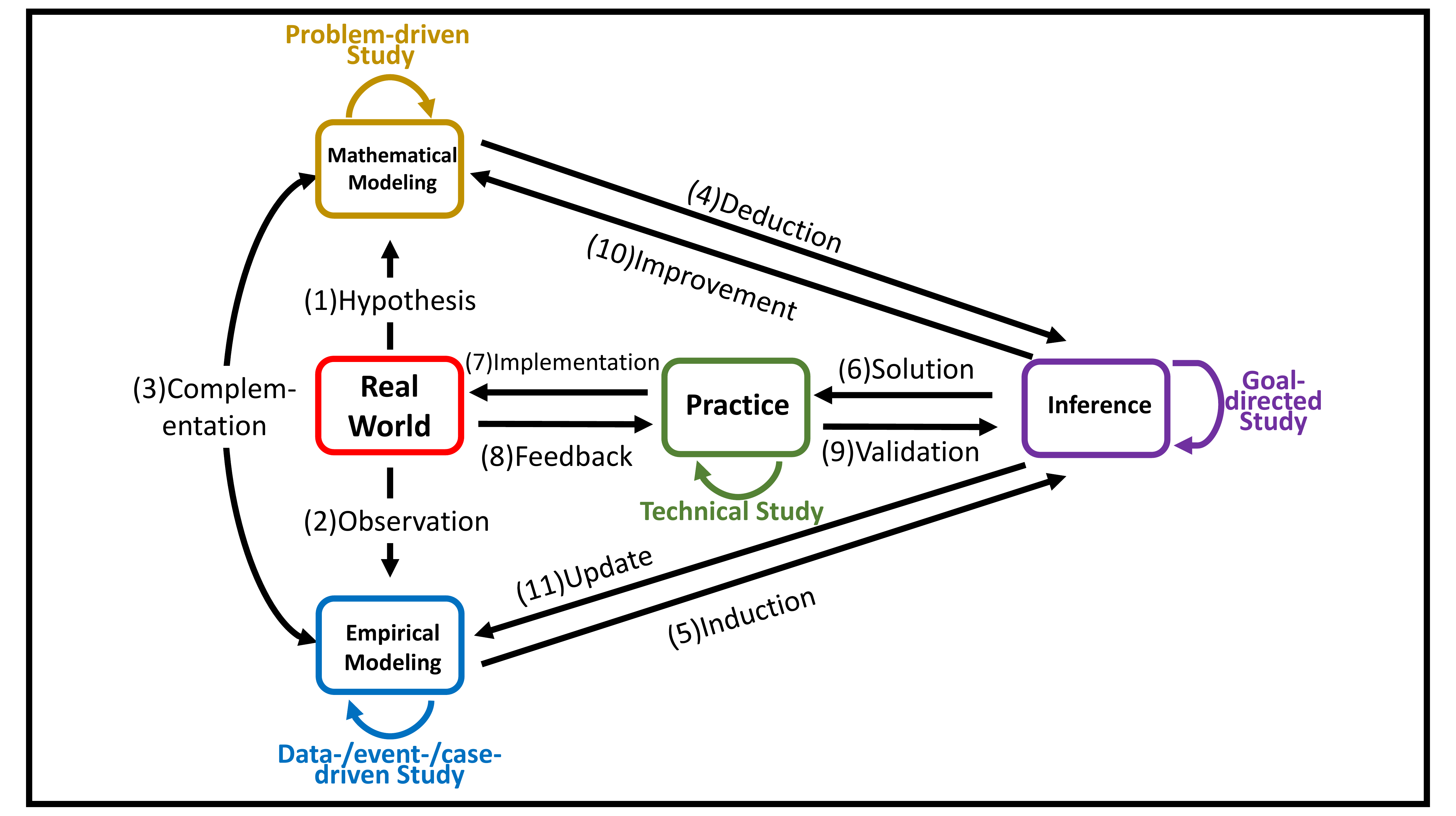}  
 \end{minipage}
 \centering
 \caption{An illustrative example of systems thinking approach for cybersecurity modeling}
\end{figure*}

\subsection{Systems Theories and Methods}
Cybersecurity modeling is a scientific way to make the cybersecurity and its related activities to represent, define, quantify and understand easier. For one proper research, in our view, the model is as equally significant as the experimentation and results analysis \cite{veksler2018simulations}. Thus, one of the difficulties in systems thinking for cybersecurity modeling is which theories and scientific methods should be most applicable in the cybersecurity model, with respective with different research scenarios and purposes. Systems thinking provides a logical method to view the cybersecurity from the guidance of the systems theories, such as system theory, cybernetics and game theory, to the assistance and analysis of relevant scientific methods, such as network analysis, dynamics and agent modeling from the particular perspective. 

Table I lists a few typical theories and scientific methods often used in the cybersecurity models briefly. There are many theories in systems thinking which refer to a set of contemplative and rational type of thinking, ideas and principles from a specific perspective. Meanwhile, a wide range of scientific methods are applied to establish the system models, understand the interactions among multiple actors, analyze the phenomena, find the explanations and predict the future. Thus, the theories and methods in one proper research are needed to tackle with specific complex cybersecurity issues.

\subsection{Relevant Stakeholders in Cybersecurity}
At the center of the cybersecurity system, stakeholder involves all aspects of the cybersecurity. They are not only the ties between all sub-systems in cybersecurity, but also act the driving force of cybersecurity. One of the vital aspects in systems thinking for cybersecurity modeling revolves around who are the relevant stakeholders in cybersecurity and how these stakeholders interact.

Not all stakeholders in cyberspace required to be considered in cybersecurity modeling. Table II lists four typical stakeholder clusters. Relevant stakeholders in the cybersecurity may include: government agencies; academia, standardization information security enterprises, private sectors, infrastructures and users. Each group of stakeholders can be considered to act as the sub-system, which has its own role in cybersecurity.

\section{An Example of Systems Framework for Cybersecurity Modeling}
Currently, cybersecurity researchers usually establish the cybersecurity model based on research fields with which they are familiar. In this section, we provide a typical systems framework for cybersecurity modeling in Fig.2. This schematic framework outlines five essential elements: 

\begin{itemize}
 \item \textbf{Real World} includes both the physical and virtual aspects of both the cyberspace. It not only includes the embodiment of concepts, parameters and equations in the mathematical model, but also provides the observations for the empirical models, such as data, events and cases.
 
 \item \textbf{Mathematical Modeling} is a type of theoretical approach to translating the behavior of the cybersecurity system into exact formulations by mathematical concepts and language. The mathematical model aims to represent what is the real-world cybersecurity problem and how the cybersecurity system evolves.
 
 \item \textbf{Empirical Modeling} is a typical study approach established from observations of cybersecurity system by measuring the system outputs, such as relevant data, security event or cyberattack case. Its goals include finding out the empirical rule or characterization of the real observation, depicting the current network security situation and estimating the probabilistic future trends.
 
 \item \textbf{Inference} refers to the process of concluding by a series of analytical methods and tools. Driven by the specific issue in cybersecurity, the inference is a purposive action, which aims to find the optimal methods and solution for the real cybersecurity.
 
 \item \textbf{Practice} is a process of study, development and implementation of the real cybersecurity solution under the guidance of analytic results. This belongs to one aspect of the technical study, which aims to covert the theoretical analysis to the real cybersecurity techniques or tools, and then apply to the real cyberspace scenario.
\end{itemize}

Mathematical models and empirical models are two significant aspects of cybersecurity modeling. A mathematical model is an abstraction or simplification of a real-world cybersecurity system and scenario, and mathematical modeling is one of the processes to perform this abstraction and simplification from the real-world cybersecurity by various mathematical tools. Notably, both the mathematical modeling and its future analysis are based on the hypotheses for the basic framework of the models (Step (1)) \cite{herley2017sok}. Then, one of the most challenges in this modeling is to find an appropriate mathematical language to establish this framework, including the mathematical equation, variables, function and so on \cite{spring2017practicing}. Empirical modeling mainly depends on empirical data in cybersecurity, such as security events, cyberattack cases, experiment results, observed and obtained from the real-world cybersecurity (Step(2)) \cite{herley2018science}. Without the specific theory and mathematical equation, this model is challenging to adopt the theoretical analysis. However, the hypothesized laws and equations in mathematical models describe the idealized system-level or network-level situations and often fail to apply to the complex cyber-level situation. The empirical model is considered as the highly feasible approach in cybersecurity modeling. Although these two modeling approaches seems different owing to the different perspective, they can complement each other (Step(3)). The empirical data are the significant source in mathematical modeling, while the mathematical model, in turn, can help refine the empirical model. 

The mathematical modeling and empirical modeling help to translate the complex cybersecurity environment into a descriptive model, which is easy for researchers to define, understand and infer by relevant knowledge. Next, researchers need to analyze the cybersecurity models further and explore the solutions to the specific cybersecurity issue. Deduction (Step (4)) always begins with the assumptions, axioms and equations in the mathematical models. The conclusion from deduction follows with certainty if the premise meets the observations of the real-world. By contrast, the inductive inference (Step (5)) is directly derived from empirical observations of the real world. For example, the efficient defense measure to the current typical cyberattack is no guarantee against encountering the new one \cite{spring2017practicing}. Based on the above inferences, the solutions or tools to the specific cybersecurity issue (Step (6)) are proposed and then implemented to tackle with the real-world cybersecurity (Step (7)).

As mentioned in Section II, it is too difficult to predict the effects of the solution from the analytic inference and its implementation because the real-world cybersecurity is complex and subtle. Therefore, an entire cybersecurity model needs the feedback from cyberspace (Step (8)) \cite{adams2013application}, adds the validation to the analytic inference method (Step (9)) and finally provides the improvement to both the mathematical and empirical modeling (Step (10) and (11)). The cybersecurity model and its analytic inference can help guide the practice in cyberspace, in turn, the feedback from the cybersecurity practice can verify the validity of the analysis and improve the current models.

\section{Conclusion}
Systems thinking allows us to think about cybersecurity modeling in a holistic and rational perspective. On the one hand, systems thinking provides a conceptual blueprint or framework for cybersecurity, where the components, factors and environments are integrated dynamically. The cybersecurity modeling with the systems approach helps us better understand and characterize the cybersecurity issues, such as unpredictability, complexity, emergence, asymmetry and dynamics, which are often ignored in most the current cybersecurity study. On other hand, through a set of analytic methods and real tools in modeling, inference and practice, systems thinking offers an innovative and universe roadmap to solve the specific cybersecurity problems, so that we can not only obtain the theoretical conclusion and the corresponding real-world solutions, but also validate the analytic conclusion and then improve the theoretical model. In this paper, despite we highlight that systems thinking should be the necessary foundation for cybersecurity modeling, the cybersecurity modeling with systems thinking still has a long way to go.

\bibliographystyle{IEEEtran}
\bibliography{IEEEabrv,ref}

\begin{thebibliography}{10}
\providecommand{\url}[1]{#1}
\csname url@samestyle\endcsname
\providecommand{\newblock}{\relax}
\providecommand{\bibinfo}[2]{#2}
\providecommand{\BIBentrySTDinterwordspacing}{\spaceskip=0pt\relax}
\providecommand{\BIBentryALTinterwordstretchfactor}{4}
\providecommand{\BIBentryALTinterwordspacing}{\spaceskip=\fontdimen2\font plus
\BIBentryALTinterwordstretchfactor\fontdimen3\font minus
  \fontdimen4\font\relax}
\providecommand{\BIBforeignlanguage}[2]{{%
\expandafter\ifx\csname l@#1\endcsname\relax
\typeout{** WARNING: IEEEtran.bst: No hyphenation pattern has been}%
\typeout{** loaded for the language `#1'. Using the pattern for}%
\typeout{** the default language instead.}%
\else
\language=\csname l@#1\endcsname
\fi
#2}}
\providecommand{\BIBdecl}{\relax}
\BIBdecl

\bibitem{craigen2014defining}
D.~Craigen, N.~Diakun-Thibault, and R.~Purse, ``Defining cybersecurity,''
  \emph{Technology Innovation Management Review}, vol.~4, no.~10, 2014.

\bibitem{al2016cyber}
H.~Al-Mohannadi, Q.~Mirza, A.~Namanya, I.~Awan, A.~Cullen, and J.~Disso,
  ``Cyber-attack modeling analysis techniques: An overview,'' in \emph{2016
  IEEE 4th International Conference on Future Internet of Things and Cloud
  Workshops (FiCloudW)}.\hskip 1em plus 0.5em minus 0.4em\relax IEEE, 2016, pp.
  69--76.

\bibitem{koutsoukos2017sure}
X.~Koutsoukos, G.~Karsai, A.~Laszka, H.~Neema, B.~Potteiger, P.~Volgyesi,
  Y.~Vorobeychik, and J.~Sztipanovits, ``Sure: A modeling and simulation
  integration platform for evaluation of secure and resilient cyber--physical
  systems,'' \emph{Proceedings of the IEEE}, vol. 106, no.~1, pp. 93--112,
  2017.

\bibitem{xi2015improved}
R.-R. Xi, X.-C. Yun, Y.-Z. Zhang, and Z.-Y. Hao, ``An improved quantitative
  evaluation method for network security,'' \emph{Chinese Journal of
  Computers}, vol.~38, no.~4, pp. 749--758, 2015.

\bibitem{herley2017sok}
C.~Herley and P.~C. Van~Oorschot, ``Sok: Science, security and the elusive goal
  of security as a scientific pursuit,'' in \emph{2017 IEEE Symposium on
  Security and Privacy (SP)}.\hskip 1em plus 0.5em minus 0.4em\relax IEEE,
  2017, pp. 99--120.

\bibitem{young2013systems}
W.~Young and N.~Leveson, ``Systems thinking for safety and security,'' in
  \emph{Proceedings of the 29th Annual Computer Security Applications
  Conference}.\hskip 1em plus 0.5em minus 0.4em\relax ACM, 2013, pp. 1--8.

\bibitem{salim2014cyber}
H.~M. Salim, ``Cyber safety: A systems thinking and systems theory approach to
  managing cyber security risks,'' Ph.D. dissertation, Massachusetts Institute
  of Technology, 2014.

\bibitem{national2017foundational}
E.~National Academies~of Sciences, Medicine \emph{et~al.}, \emph{Foundational
  Cybersecurity Research: Improving Science, Engineering, and
  Institutions}.\hskip 1em plus 0.5em minus 0.4em\relax National Academies
  Press, 2017.

\bibitem{spring2017practicing}
J.~M. Spring, T.~Moore, and D.~Pym, ``Practicing a science of security: a
  philosophy of science perspective,'' in \emph{Proceedings of the 2017 New
  Security Paradigms Workshop}.\hskip 1em plus 0.5em minus 0.4em\relax ACM,
  2017, pp. 1--18.

\bibitem{phister2011cyberspace}
P.~W. Phister~Jr, ``Cyberspace: The ultimate complex adaptive system,'' OFFICE
  OF THE ASSISTANT SECRETARY OF DEFENSE WASHINGTON DC COMMAND AND~…, Tech.
  Rep., 2011.

\bibitem{rzeszutko2015insights}
E.~Rzeszutko and W.~Mazurczyk, ``Insights from nature for cybersecurity,''
  \emph{Health security}, vol.~13, no.~2, pp. 82--87, 2015.

\bibitem{dunlavy2009mathematical}
D.~M. Dunlavy, B.~Hendrickson, and T.~G. Kolda, ``Mathematical challenges in
  cybersecurity,'' \emph{Sandia Report, February}, 2009.

\bibitem{armstrong2009complexity}
R.~Armstrong, J.~Mayo, and F.~Siebenlist, ``Complexity science challenges in
  cybersecurity,'' \emph{Sandia National Laboratories SAND Report}, 2009.

\bibitem{oltramari2015towards}
A.~Oltramari, D.~S. Henshel, M.~Cains, and B.~Hoffman, ``Towards a human
  factors ontology for cyber security.'' in \emph{STIDS}, 2015, pp. 26--33.

\bibitem{xu2015emergent}
S.~Xu, ``Emergent behavior in cybersecurity,'' \emph{arXiv preprint
  arXiv:1502.05102}, 2015.

\bibitem{xu2015cybersecurity}
------, ``Cybersecurity dynamics,'' \emph{arXiv preprint arXiv:1502.05100},
  2015.

\bibitem{yan2019dynamical}
D.~Yan, F.~Liu, Y.~Zhang, and K.~Jia, ``Dynamical model for individual defence
  against cyber epidemic attacks,'' \emph{Iet Information Security}, vol.~13,
  no.~6, pp. 541--551, 2019.

\bibitem{geers2010challenge}
K.~Geers, ``The challenge of cyber attack deterrence,'' \emph{Computer Law \&
  Security Review}, vol.~26, no.~3, pp. 298--303, 2010.

\bibitem{Cai2016Moving}
G.~L. Cai, B.~S. Wang, H.~U. Wei, and T.~Z. Wang, ``Moving target defense:state
  of the art and characteristics,'' \emph{Frontiers of Information Technology
  and Electronic Engineering}, vol.~17, no.~11, pp. 1122--1153, 2016.

\bibitem{tisdale2015cybersecurity}
S.~M. Tisdale, ``Cybersecurity: Challenges from a systems, complexity,
  knowledge management and business intelligence perspective.'' \emph{Issues in
  Information Systems}, vol.~16, no.~3, 2015.

\bibitem{manshaei2013game}
M.~H. Manshaei, Q.~Zhu, T.~Alpcan, T.~Bac{\c{s}}ar, and J.-P. Hubaux, ``Game
  theory meets network security and privacy,'' \emph{ACM Computing Surveys
  (CSUR)}, vol.~45, no.~3, p.~25, 2013.

\bibitem{do2017game}
C.~T. Do, N.~H. Tran, C.~Hong, C.~A. Kamhoua, K.~A. Kwiat, E.~Blasch, S.~Ren,
  N.~Pissinou, and S.~S. Iyengar, ``Game theory for cyber security and
  privacy,'' \emph{ACM Computing Surveys (CSUR)}, vol.~50, no.~2, p.~30, 2017.

\bibitem{adams2013application}
M.~D. Adams, S.~D. Hitefield, B.~Hoy, M.~C. Fowler, and T.~C. Clancy,
  ``Application of cybernetics and control theory for a new paradigm in
  cybersecurity,'' \emph{arXiv preprint arXiv:1311.0257}, 2013.

\bibitem{vinnakota2013cybernetics}
T.~Vinnakota, ``A cybernetics paradigms framework for cyberspace: Key lens to
  cybersecurity,'' in \emph{2013 IEEE International Conference on Computational
  Intelligence and Cybernetics (CYBERNETICSCOM)}.\hskip 1em plus 0.5em minus
  0.4em\relax IEEE, 2013, pp. 85--91.

\bibitem{wiederhold2014role}
B.~K. Wiederhold, ``The role of psychology in enhancing cybersecurity,'' 2014.

\bibitem{marble2015human}
J.~L. Marble, W.~F. Lawless, R.~Mittu, J.~Coyne, M.~Abramson, and C.~Sibley,
  ``The human factor in cybersecurity: Robust \& intelligent defense,'' in
  \emph{Cyber Warfare}.\hskip 1em plus 0.5em minus 0.4em\relax Springer, 2015,
  pp. 173--206.

\bibitem{anwar2017gender}
M.~Anwar, W.~He, I.~Ash, X.~Yuan, L.~Li, and L.~Xu, ``Gender difference and
  employees' cybersecurity behaviors,'' \emph{Computers in Human Behavior},
  vol.~69, pp. 437--443, 2017.

\bibitem{zheng2015active}
R.~Zheng, W.~Lu, and S.~Xu, ``Active cyber defense dynamics exhibiting rich
  phenomena,'' in \emph{Proceedings of the 2015 Symposium and Bootcamp on the
  Science of Security}.\hskip 1em plus 0.5em minus 0.4em\relax ACM, 2015, p.~2.

\bibitem{la2014role}
R.~J. La, ``Role of network topology in cybersecurity,'' in \emph{53rd IEEE
  Conference on Decision and Control}.\hskip 1em plus 0.5em minus 0.4em\relax
  IEEE, 2014, pp. 5290--5295.

\bibitem{pino2014network}
R.~E. Pino, \emph{Network science and cybersecurity}.\hskip 1em plus 0.5em
  minus 0.4em\relax Springer, 2014.

\bibitem{such2016intelligent}
J.~M. Such, N.~Criado, L.~Vercouter, and M.~Rehak, ``Intelligent cybersecurity
  agents [guest editors' introduction],'' \emph{IEEE Intelligent Systems},
  vol.~31, no.~5, pp. 3--7, 2016.

\bibitem{white2017computer}
G.~B. White, E.~A. Fisch, and U.~W. Pooch, \emph{Computer system and network
  security}.\hskip 1em plus 0.5em minus 0.4em\relax CRC press, 2017.

\bibitem{stallings2017cryptography}
W.~Stallings, \emph{Cryptography and network security: principles and
  practice}.\hskip 1em plus 0.5em minus 0.4em\relax Pearson Upper Saddle River,
  2017.

\bibitem{enterprise2018}
Verizon. (2018) 2018 data breach investigations report.

\bibitem{yan2018characterizing}
D.~Yan, F.~Liu, Y.~Zhang, K.~Jia, and Y.~Zhang, ``Characterizing the optimal
  attack strategy decision in cyber epidemic attacks with limited resources,''
  in \emph{International Conference on Science of Cyber Security}.\hskip 1em
  plus 0.5em minus 0.4em\relax Springer, 2018, pp. 65--80.

\bibitem{mclaughlin2016cybersecurity}
S.~McLaughlin, C.~Konstantinou, X.~Wang, L.~Davi, A.-R. Sadeghi, M.~Maniatakos,
  and R.~Karri, ``The cybersecurity landscape in industrial control systems,''
  \emph{Proceedings of the IEEE}, vol. 104, no.~5, pp. 1039--1057, 2016.

\bibitem{yan2019modeling}
D.~Yan, F.~Liu, and K.~Jia, ``Modeling an information-based advanced persistent
  threat attack on the internal network,'' in \emph{ICC 2019-2019 IEEE
  International Conference on Communications (ICC)}.\hskip 1em plus 0.5em minus
  0.4em\relax IEEE, 2019, pp. 1--7.

\bibitem{ask2013advanced}
M.~Ask, P.~Bondarenko, J.~E. Rekdal, A.~Nordb{\o}, P.~Bloemerus, and
  D.~Piatkivskyi, ``Advanced persistent threat (apt) beyond the hype,''
  \emph{Project Report in IMT4582 Network Security at GjoviN University
  College}, 2013.

\bibitem{Liang2016The}
G.~Liang, S.~R. Weller, J.~Zhao, F.~Luo, and Z.~Y. Dong, ``The 2015 ukraine
  blackout: Implications for false data injection attacks,'' \emph{IEEE
  Transactions on Power Systems}, vol.~PP, no.~99, pp. 1--1, 2016.

\bibitem{Pendleton2016A}
M.~Pendleton, R.~Garcia-Lebron, J.~H. Cho, and S.~Xu, ``A survey on systems
  security metrics,'' \emph{Acm Computing Surveys}, vol.~49, no.~4, p.~62,
  2016.

\bibitem{arnold2015definition}
R.~D. Arnold and J.~P. Wade, ``A definition of systems thinking: a systems
  approach,'' \emph{Procedia Computer Science}, vol.~44, pp. 669--678, 2015.

\bibitem{de2009systems}
D.~De~Savigny and T.~Adam, \emph{Systems thinking for health systems
  strengthening}.\hskip 1em plus 0.5em minus 0.4em\relax World Health
  Organization, 2009.

\bibitem{xia2017systems}
S.~Xia, X.-N. Zhou, and J.~Liu, ``Systems thinking in combating infectious
  diseases,'' \emph{Infectious diseases of poverty}, vol.~6, no.~1, p. 144,
  2017.

\bibitem{veksler2018simulations}
V.~D. Veksler, N.~Buchler, B.~E. Hoffman, D.~N. Cassenti, C.~Sample, and
  S.~Sugrim, ``Simulations in cybersecurity: A review of cognitive modeling of
  network attackers, defenders, and users,'' \emph{Frontiers in psychology},
  vol.~9, p. 691, 2018.

\bibitem{herley2018science}
C.~Herley and P.~C. Van~Oorschot, ``Science of security: Combining theory and
  measurement to reflect the observable,'' \emph{IEEE Security \& Privacy},
  vol.~16, no.~1, pp. 12--22, 2018.

\end{thebibliography}

\end{document}